\documentclass[12pt]{article}
\usepackage{amssymb,amsmath}
\usepackage[noblocks]{authblk}
\usepackage[top=0.75in, bottom=0.75in, left=0.75in, right=0.75in, dvips]{geometry}
\usepackage{caption}  
\date{}   

\begin{document}
\textwidth 10.0in       
\textheight 9.0in 
\topmargin -0.60in
\title{A Gauge Theory that Mixes Bosonic and\\
Fermionic Gauge Fields}
\author[]{D.G.C. McKeon\thanks{Email: dgmckeo2@uwo.ca}}
\affil[] {Department of Applied Mathematics, The
University of Western Ontario, London, ON N6A 5B7, Canada}
\affil[] {Department of Mathematics and
Computer Science, Algoma University, Sault Ste. Marie, ON P6A
2G4, Canada}  
\maketitle    

\maketitle
\noindent
PACS No.:  11.30 Pb\\
Key Words: gauge theory

\begin{abstract}
Using a gauge symmetry derived by applying the Dirac constraint formalism to supergravity with a cosmological term in $2+1$ dimensions, we construct a gauge theory with many characteristics of Yang-Mills theory.  The gauge transformation mixes two Bosonic fields and one Fermionic field.
\end{abstract}

In $2+1$ dimensions, the Einstein-Cartan action takes the form of a topological field theory [1,2].
\begin{equation}
S_1 = \int d^3x \epsilon^{\mu\nu\lambda} b_\mu^i R_{\nu\lambda i}
\end{equation}
where
\begin{equation}
R_{\nu\lambda i} = \partial_\mu w_{\nu i} - \partial_\nu w_{\mu i} - \epsilon_{ijk} w_\mu^j  w_\nu^k
\end{equation}
with $b_\mu^i$ and $w_\mu^i$ being independent fields.

We can supplement $S_1$ with a term $S_2$ that involves a Majorana field $\psi_\mu$ [3]
\begin{equation}
S_2 = \int d^3x \epsilon^{\mu\nu\lambda} \overline{\psi}_\mu D_\nu \psi_\lambda
\end{equation}
where
\begin{equation}
D_\mu = \partial_\mu + \frac{i}{2} \gamma^i w_{\mu i}.
\end{equation}
The application of the Dirac constraint analysis to $S_1 + S_2$ results in a set of first and second class constraints; when the method of Henneaux-Teitelboim and Zanelli (HTZ) [4] is then applied, it is found that $S_1 + S_2$ is invariant under a set of gauge transformations that mix $b_\mu^i$, $w_\mu^i$ and $\psi_\mu$ that is distinct from the usual supergravity gauge transformations [5].  These transformations do not require introduction of auxiliary fields or imposition of the equations of motion in order to close.

We now can supplement $S_1 + S_2$ with ``cosmological constant'' terms
\begin{equation}
S_K = \int d^3x \epsilon^{\mu\nu\lambda} \left( \frac{\Lambda}{3}\epsilon_{ijk} b_\mu^i b_\nu^j b_\lambda^k + \frac{iK}{2} \overline{\psi}_\mu b_\nu^i \gamma_i \psi_\lambda \right).
\end{equation}
In ref. [6] the Dirac constraint analyse when supplemented by the HTZ procedure shows that $S = S_1 + S_2 + S_K$ is invariant
under the gauge transformation
\begin{subequations}
\begin{align}
\delta b_{i\mu} &= \partial_\mu A_i + \epsilon_{ipj} B^p b_\mu^j + \epsilon_{ipj} A^p w_\mu^j + \frac{i}{2}\overline{C} \gamma_i \psi_\mu\\
\delta w_{i\mu} &= \partial_\mu B_i + K^2\epsilon_{ipj} A^p b_\mu^j + \epsilon_{ipj} B^p w_\mu^j + \frac{iK}{2}\overline{C} \gamma_i \psi_\mu\\
\delta \psi_\mu &= \partial_\mu C + \frac{iK}{2} \gamma_j b_\mu^j C + \frac{i}{2} \gamma_j w_\mu^j C - \frac{i}{2}\gamma_j
(K \;A^j + B^j) \psi_\mu
\end{align}
\end{subequations}
provided that in eq. (5)
\begin{equation}
\Lambda = - K^2.
\end{equation}
(A misprint in ref. [6] is corrected in eq. (6).)  In eq. (6), $A^i$ and $B^i$ are Bosonic gauge functions and $C$ is a Fermionic Majorana spinor gauge function.

If $(A_1^i, B_1^i, C_1)$ and $(A_2^i, B_2^i, C_2)$ are gauge functions associated with gauge transformations $\delta_1$ and $\delta_2$, then from eq. (6) it follows that $\delta_1\delta_2 - \delta_2\delta_1$ is itself a gauge transformation of the form of eq. (6) with gauge functions
\begin{subequations}
\begin{align}
\tilde{A}_i &= \epsilon_{ijk} \left(B_1^j A_2^k + A_1^j B_2^k\right) + \frac{i}{2} \overline{C}_1 \gamma_i C_2\\
\tilde{B}_i &= \epsilon_{ijk} \left(K^2 A_1^j A_2^k + B_1^j B_2^k\right) + \frac{iK}{2} \overline{C}_1 \gamma_i C_2\\
\tilde{C} &= \frac{iK}{2} A_2^j \gamma_j C_1 + \frac{i}{2} B_2^j \gamma_j C_1 - \frac{iK}{2} A_1^j \gamma_j C_2 -
\frac{i}{2}B_1^j \gamma_j C_2
\end{align}
\end{subequations}
in place of $A_i$, $B_i$ and $C$.

If now we were to write eq. (6) in the form
\begin{equation}
\delta \Phi_\mu = \partial_\mu \Theta + \underset{\sim}{V}_\mu
 (\Phi)\Theta
\end{equation}
where
\begin{subequations}
\begin{align}
\Phi_\mu &= \left( \begin{array}{c}
b^i_\mu \\ w_\mu^i \\ \psi_\mu \end{array} \right) \\
\Theta &= \left( \begin{array}{c}
A^i \\ B^i \\ C \end{array} \right)
\end{align}
\end{subequations}
and
\begin{equation}
\underset{\sim}{V}_\mu (\Phi) = \left(
\begin{array}{lll}
-\epsilon_{ipj}w_\mu^p & -\epsilon_{ipj} b_\mu^p & -\frac{i}{2} \overline{\psi}_\mu \gamma_i\\
-K^2\epsilon_{ipj} b_\mu^p & -\epsilon_{ipj}w_\mu^p & -\frac{iK}{2} \overline{\psi}_\mu \gamma_i\\
-\frac{iK}{2} \gamma_j\psi_\mu &  -\frac{i}{2} \gamma_j\psi_\mu & \frac{i}{2} \gamma_p \left(Kb_\mu^p + w_\mu^p\right)
\end{array} \right)
\end{equation}
then eq. (9) has the form of gauge transformation in Yang-Mills theory. To see this, we write eq. (6) as
\begin{equation}
\delta \Phi_\mu = \partial_\mu \Theta + \underset{\sim}{\Omega} (\Theta) \Phi_\mu
\end{equation}
where
\begin{equation}
\underset{\sim}{\Omega} (\Theta) = \left(
\begin{array}{lll}
\epsilon_{ipj}B^p & \epsilon_{ipj} A^p & \frac{i}{2} \overline{C} \gamma_i\\
K^2\epsilon_{ipj}A^p & \epsilon_{ipj}B^p & \frac{iK}{2} \overline{C} \gamma_i\\
\frac{iK}{2} \gamma_j C &  \frac{i}{2} \gamma_j C & -\frac{i}{2} \gamma_p \left(K A^p + B^p\right)
\end{array} \right)
\end{equation}
and then note that eq. (6) can be written as
\begin{equation}
\delta \underset{\sim}{V}_\mu (\Phi) = - \partial_\mu \underset{\sim}{\Omega}(\Theta) - \left[ \underset{\sim}{\Omega}, 
(\Theta), \underset{\sim}{V}_\mu (\Phi)\right].
\end{equation}
Eq. (14) has the form of a gauge transformation in Yang-Mills theory and so we can define a ``field strength''
\begin{equation}
\mathcal{\underset{\sim}{F}}_{\mu\nu} (\Phi) = \partial_\mu \underset{\sim}{V}_\nu (\Phi) - \partial_\nu \underset{\sim}{V}_\mu (\Phi) +
\left[  \underset{\sim}{V}_\mu (\Phi),\underset{\sim}{V}_\nu (\Phi) \right]
\end{equation}
which under the gauge transformation of eq. (6) transforms as
\begin{equation}
\delta \mathcal{\underset{\sim}{F}}_{\mu\nu} = - \left[ \underset{\sim}{\Omega}, \mathcal{\underset{\sim}{F}}_{\mu\nu}\right].
\end{equation}
From eqs. (11,15) we see that the entries in $\mathcal{F}_{\mu\nu}$ are
\begin{subequations}
\begin{align}
\left[ \left(\mathcal{F}_{\mu\nu}\right)_{ij}\right]_{11} &= -\epsilon_{ipj} \left(\partial_\mu w_\nu^p - \partial_\nu w_\mu^p\right) + \left(w_{j\mu} w_{i\nu} - w_{j\nu}w_{i\mu}\right)\\
& + K^2 \left(b_{j\mu} b_{i\nu} - b_{j\nu} b_{i\mu}\right) - \frac{i}{2}\epsilon_{ijk} \overline{\psi}_\mu \gamma^k\psi_\nu\nonumber\\
&= \left[ \left(\mathcal{F}_{\mu\nu}\right)_{ij}\right]_{22}\\
\left[ \left(\mathcal{F}_{\mu\nu}\right)\right]_{33} &= \frac{i\gamma_p}{2}\left[K\left(\partial_\mu b_\nu^p - \partial_\nu b_\mu^p\right) + \left(\partial_\mu w_\nu^p - \partial_\nu w_\mu^p\right)\right]\\
& + \left[ \frac{K}{2} \overline{\psi}_\mu \gamma_p\psi_\mu - \frac{i}{2} \epsilon_{mnp} 
\left( K b_\mu^m + w_\mu^m\right)\left(K b_\nu^n + w_\nu^n\right)\right]\gamma^p\nonumber\\
\left[ \left(\mathcal{F}_{\mu\nu}\right)_{ij}\right]_{12} &= -\epsilon_{ipj} \left( \partial_\mu b_\nu^p - \partial_\nu b_\mu^p\right) + 
\left( w_{j\mu} b_{i\nu} - w_{j\nu}b_{i\mu}\right)\\
&+ \left( b_{j\mu} w_{i\nu} - b_{j\nu} w_{i\mu} \right) - \frac{i}{2} \epsilon_{ijk} \overline{\psi}_\mu \gamma^k \psi_\nu\nonumber \\
&\left[ \left(\mathcal{F}_{\mu\nu}\right)_{ij}\right]_{21} = K^2 \left[ \left(\mathcal{F}_{\mu\nu}\right)_{ij}\right]_{12} \\
\left[ \left(\mathcal{F}_{\mu\nu}\right)_i\right]_{13} &= -\frac{i}{2} \left(\partial_\mu \overline{\psi}_\nu - \partial_\nu\overline{\psi}_\mu\right)\gamma_i +
\frac{1}{4}\big[ \overline{\psi}_\mu \left(K b_{i\nu} + w_{i\nu}\right) - \overline{\psi}_\nu\left(K b_{i\mu} + w_{i\mu}\right)\\
&+ i\epsilon_{ipq} \left( \overline{\psi}_\nu \left( K b_\mu^p + w_\mu^p \right) - \overline{\psi}_\mu\left(K b_\nu^p + w_\nu^p\right)\right)\gamma^q \big]\nonumber \\
\left[ \left(\mathcal{F}_{\mu\nu}\right)_j\right]_{32} &= -\frac{i}{2} \gamma_j\left(\partial_\mu \psi_\nu - \partial_\nu\psi_\mu\right)+
\frac{K}{4}\big[ \left(K b_{j\mu} + w_{j\mu}\right) \psi_\nu - \left(K b_{j\nu} + w_{j\nu}\right)\psi_\mu\\
&+ i\epsilon_{pqj} \gamma^p \left(\left( K b_\nu^q + w_\nu^q \right) \psi_\mu - \left(K b_\mu^q + w_\mu^q\right)\psi_\nu\right)\big]\nonumber\\
&\left[ \left(\mathcal{F}_{\mu\nu}\right)_j\right]_{31} = K\left[ \left(\mathcal{F}_{\mu\nu}\right)_j\right]_{32}\\
&\left[ \left(\mathcal{F}_{\mu\nu}\right)_i\right]_{23} = K\left[ \left(\mathcal{F}_{\mu\nu}\right)_i\right]_{13} .
\end{align}
\end{subequations}

An action that is invariant under the gauge transformation of eq. (6) (or alternatively, eq. (12)) is
\begin{equation}
S_\Phi = -\frac{1}{4} \int d^nx Str \left( \mathcal{\underset{\sim}{F}}_{\mu\nu} (\Phi) \mathcal{\underset{\sim}{F}}^{\mu\nu} (\Phi) \right).
\end{equation}
If eq. (17) is used to express $S_\Phi$ in terms of the two vectors $b_\mu^i$, $w_\mu^i$ and the spinor $\psi_\mu$, one is left with an  exceedingly complicated expression that would make calculation of radiative effects quite difficult.  An alternative is to use a first order form for $S_\Phi$,
\begin{equation}
\tilde{S}_\Phi = \int d^nx \left( - \frac{1}{2} Str\left( \mathcal{\underset{\sim}{F}}_{\mu\nu} (\Phi) \underset{\sim}{F}^{\mu\nu}\right) + \frac{1}{4}
Str \left( \underset{\sim}{F}_{\mu\nu}\underset{\sim}{F}^{\mu\nu}\right) \right)
\end{equation}
where $\underset{\sim}{F}_{\mu\nu}$ and $\Phi_\mu$ are treated as independent fields.  In eq. (19), all vertices are at most cubic in the fields and independent of momenta.  The use of such a first order Lagrangian also simplifies calculations in Yang-Mills theory [7].

In eqs. (1,3,5) it is evident that the Greek indices $\mu$, $\nu$, $\lambda$ are vector indices in a $2 + 1$ dimensional space.  However, in eq. (18), these indices need not be so restricted; they can be in an $n$-dimensional space (including $3 + 1$ dimensions).

However, the Latin indices $i$, $j$, etc. are still restricted to being in $2 + 1$ dimensions with the space $\psi_\mu$ being a two component Majorana spinor.  It would be interesting to see if this restriction could be relaxed.  Going from $2 + 1$ to $3 + 0$ dimensions does not appear to be feasible as it is not possible to have a Majorana spinor in $3 + 0$ dimensions, and having two symplectic Majorana spinors satisfying $\psi_{1_\mu} = \psi_{2c_\mu}, \psi_{2_\mu} = -\psi_{1c_\mu}$ does not appear to be compatible with any modification of eq. (6) that is consistent with an analogue of eq. (8).  However, it is possible to consider the case in which the Fermionic field $\psi_\mu$ as well as the Fermionic gauge function $C$ vanish.  In doing so, we have a purely Bosonic gauge theory with two gauge fields $w_\mu^i$ and $b_\mu^i$.  These can be taken to be in a space with metric $(+ + +)$.

It would also be of interest to couple matter fields to the gauge fields $\Phi_\mu$ and to determine if divergences arising in radiative effects could be removed through renormalization.

\section*{Acknowledgements}

Roger Macleod initiated this research.

\section*{Appendix-Notation}

We use the metric $\eta^{00} = -\eta^{11} = - \eta^{22} = 1$, and have $\epsilon^{012} = 1$.  Dirac matrices are imaginary, $\gamma^0 = \sigma_2$, $\gamma^1 = i\sigma_3$, $\gamma^2 = i\sigma$, with
\begin{equation}\tag{A.1}
\gamma^1 \gamma^j = \eta^{ij} + i\epsilon^{ijk} \gamma_k
\end{equation}
and
\begin{equation}\tag{A.2}
(\gamma^i)_{ab} (\gamma_i)_{cd} = -\frac{1}{2} (\gamma^i)_{ad} (\gamma_i)_{cb} + \frac{3}{2}
\delta_{ad} \delta_{cb}.
\end{equation}
We impose the Majorana condition on spinors so that 
\begin{equation}\tag{A.3}
\psi = -\gamma^0 \overline{\psi}^T = \psi^* \;\; (\overline{\psi} \equiv \psi^+ \gamma^0)
\end{equation}
which leads to 
\begin{equation}\tag{A.4}
\overline{\psi}\chi = \overline{\chi}\psi, \;\; \overline{\psi} \gamma^i \chi = - \overline{\chi}\gamma^i \psi.
\end{equation}

\end{document}